# The metrology cameras for Subaru PFS and FMOS

Shiang-Yu Wang*[a], Yen-Shan Hu[a], Chi-Hung Yan[a], Yin-Chang Chang[a], Naoyuki Tamura[b], Naruhisa Takato[c], Atsushi Shimono[b], Jennifer Karr[a], Youichi Ohyama[a], Hsin-Yo Chen[a], Hung-Hsu Ling[a], Hiroshi Karoji[b], Hajime Sugai[b], Akitoshi Ueda[b]

[a]Academia Sinica, Institute of Astronomy and Astrophysics, P. O. Box 23-141, Taipei, Taiwan; [b]Kavli Institute for the Physics and Mathematics of the Universe, the University of Tokyo, 5-1-5 Kashiwanoha, Kashiwa, 277-8583, Japan; [c]Subaru Telescope, National Astronomical Observatory of Japan, 650 North A'ohoku Place Hilo, HI 96720, U.S.A.

## ABSTRACT

The Prime Focus Spectrograph (PFS) is a new multi-fiber spectrograph on Subaru telescope. PFS will cover around 1.4 degree diameter field with ~2400 fibers. To ensure precise positioning of the fibers, a metrology camera is designed to provide the fiber position information within 5 μm error. The final positioning accuracy of PFS is targeted to be less than 10 μm. The metrology camera will locate at the Cassegrain focus of Subaru telescope to cover the whole focal plan. The PFS metrology camera will also serve for the existing multi-fiber infrared spectrograph FMOS.

**Keywords:** Metrology, CMOS sensor, multi-fiber, spectrograph

## 1. INTRODUCTION

The Prime Focus Spectrograph (PFS) is a new multi-fiber spectrograph on Subaru telescope. PFS will provide low resolution spectrum for the scientific objects from 390nm to 1.3μm. PFS shares the same wide field corrector (WFC) with Hyper SuprimeCam[1] which is a new wide field camera with 1.5 degree field of view. The 2400 fibers populate in a hexagon shape on the focal plane of PFS covering roughly 1.3~1.4 degree diameter field. Every fiber is designed to be driven by the COBRA module which has two miniature motors to provide two degree of freedom in 9mm diameter region on the focal plane. The function of the metrology camera is to measure the location of the fiber tips as a feedback signal of positioning for the COBRA system. The metrology camera produces an image of the back-lit fiber tips on the positioner focal plane. It then calculates the positions of these images on the metrology camera coordinates system for COBRA configuration system to convert these into fiber positions on the focal plane. With several iterations of COBRA movements and metrology camera measurements, the fibers can be accurately positioned on the selected science targets.

Based on the scientific requirement of positioning the fibers within 0.1" (or 10 μm) to the science targets, the metrology camera shall measure the in-plane position of each fiber tip within an error of 5 μm with a goal of 3μm. To reduce the overhead in fiber positioning, the metrology camera shall measure the centroid of all science fibers in less than 5s. As the PFS camera is capable of providing good image quality through WFC of FMOS[2], the Subaru infrared multi-fiber spectrograph, it will also be used as the metrology system for FMOS fiber positioner. Currently, the configuration of FMOS fibers is done by the focal plane scanning camera. The configuration time is typically about 15 minutes. With the metrology camera, the overhead can be much reduced.

The PFS collaboration is led by Kavli Institute for the Physics and Mathematics of the Universe, the University of Tokyo with international partners consisting of Universidade de São Paulo/Laboratório Nacional de Astrofísica in Brazil, Caltech/Jet Propulsion Laboratory, Princeton University/John Hopkins University in USA, Laboratoire d'Astrophysique de Marseille in France, Academia Sinica, Institute of Astronomy and Astrophysics in Taiwan, and National Astronomical Observatory of Japan/Subaru Telescope.

* sywang@asiaa.sinica.edu.tw; phone 886 2 2366-5338; fax 886 2 2367-7849; www.asiaa.sinica.edu.tw

## 2. METROLOGY SYSTEM ANALYSIS

### 2.1 Requirement analysis

The most critical requirement for the metrology camera is to determine the positions of the fibers on the focal plane within an error of 5 µm. The requirement affects the number of pixels of the camera, pixel size, the sub-pixel quality of the image sensor, image quality, and the illumination intensity. To fulfill the requirements, the metrology camera has to cover the whole focal plane (460mm in diameter) and provides enough spatial resolution to reach the required centroid measurement accuracy. It has been suggested that compact source image but extending over at least ~4 pixels across is needed to give 1/25 pixel centroiding accuracy[3]. With single metrology camera, the following two criteria determine the magnification factor of the metrology camera lens, M:

1. Physical size of sensor chip > M × 460mm

2. Pixel size of sensor chip/25 < M × 5µm

With the tight requirements, the detector for the metrology camera requires large number of pixels. Conventional scientific CCDs are not suitable for this. On the other hand, large format commercial CMOS sensors provide the suitable pixel size and pixel number required for the metrology camera. Typical CMOS sensors are of the size close to APS film roughly 28mm × 21mm. With this physical size, the pixel size should be less than 3.3 µm and the format should be larger than 6K × 6K to meet the criteria. Such CMOS sensors have been developed and deployed in commercial digital cameras. Furthermore, faster readout speed is offered by CMOS sensors compared with CCDs. It is advantageous since the camera needs fast readout to meet the requirement of less than 40s positioning time. With 6 to 7 iterations of COBRA move to establish the required position accuracy, the time allowed for a single exposure, readout and processing is roughly 6 second. The shorter readout time provides shorter overall overhead.

To map the locations of the fibers to the focal plane coordinate, the distortion and magnification factors of the metrology camera should be precisely measured. Considering the focal plane radius of 230mm with positioning accuracy of 5 µm, the accuracy of distortion and magnification factor should be better than $2.2 \times 10^{-5}$ at different elevation angles and telescope temperatures. This is not practical given most of the distortion is from the WFC and the combined distortion with the metrology camera optics could not be easily estimated. A set of fixed reference points on the focal plane is required to achieve the accuracy. On the focal plane, certain number of fiducial fibers will be installed as the reference points. However, in order to maximize the number of science fibers, the number of fiducial fibers should be limited. There is a tradeoff between the number of scientific fibers and the positioning accuracy defined by the density of fiducial fibers. In PFS, the home positions of the COBRA motors can be used as the reference points to calculate the distortion map of the WFC. After the COBRA moves, the fiducial fibers will be used to remove the local image motion from the dome seeing effect. It might be possible that the dome seeing is too large that we need reference points within certain distance to accurately measure the image motion at all time for any COBRA configuration.

### 2.2 Dome seeing effect

The image of fibers obtained from the metrology camera might suffer from the turbulence inside the dome of Subaru telescope. Given the tight accuracy requirement, the dome seeing effect is crucial to the best achievable precision of metrology images. The image motion generated by the dome seeing might depend on the local turbulence and impose a maximum separation of two fiducial fibers for the accuracy we need. To investigate the dome seeing at Subaru telescope, a series of movies of FMOS Echidna fiber images were taken and analyzed with different exposure time and telescope shutter condition in order to see the degree of dome seeing effect and possible dependence on dome flushing.

The FMOS fiber tips were imaged through WFC for FMOS, M3opt, folding mirror and the telescope lens of a commercial CMOS camera. The folding mirror and the camera were temporarily placed near the optical axis of the NsOpt light path, so the fiber tips were viewed slightly off-axis. The fibers were in home position. The magnification factor of the camera is about 0.0623. The camera pixel size is 2.2µm. The position of the fiber spots in the images is calculated by Sextractor with Gaussian fitting. After that, the common motion among the fiber spots was removed from the vibration of Subaru telescope and the camera mount. Only the relative image movement is discussed here.

In general, the files taken with shorter exposure time show larger relative image motion between fiber spots compared with the long exposure images. This is expected the higher single to noise ratio in the long exposure images gives better centroid calculation accuracy. Also the high frequency seeing effect might be averaged out. In most files, the relative motion between two spots decreases as the relative distance gets shorter and approaches to some lower limit when the distance is shorter than certain distance. Figure 1 shows the result with 1s frame exposure time. The relative motion decreases linearly with the distance between two spots. The smallest relative motion is smaller than 0.1 pixel which is ~3.5μm at the focal plane. The smallest relative motion is about 2μm. It is known that the camera was operated in a rolling shutter mode so the image of the spots of different rows was actually taken at different timing. This might generate extra relative motion between spots in different rows. The result is critical to the density of fiducial fibers and the minimum exposure time of metrology camera. More images are scheduled to be taken to fully explore the dome effect.

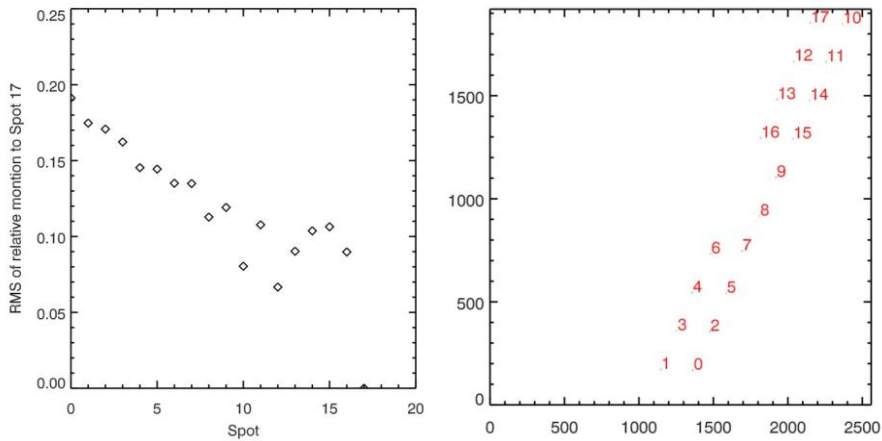

Figure 1. The relative movement of fibers respected fiber spot 17. The location for the fiber spots in the image is shown in the right panel. The relative motion stops to decrease from fiber 10~16. The distance of this region is about 700 pixels or 25mm on the focal plane.

## 3. BASELINE DESIGN

The baseline of the metrology camera design is a single CMOS camera mounted at the Cassegrain focus to directly image the fibers on the focal plane of the prime focus. The camera will be installed inside a Casegrain instrument box like other Subaru Cass instruments. Calibration of the image distortion and mapping to the focal plane through the WFC is achieved by back-lit fixed fiducial fibers and possibly by home reference positions of scientific fibers, located on the focus plane, whose positions have been previously measured to a high repeatable accuracy.

### 3.1 Camera detector

The baseline camera sensor is a 120M pixel CMOS sensor with 2.2μm pixels. The prototype camera has been developed in Canon in the spring of 2012. We expect to get an engineering device in July of 2012 for detailed tests. The physical size of the sensor is about 28mm x 21mm. It is a frontside illuminated sensor with a micro-lens on every pixel to improve the light collection efficiency. A 10M pixel version of Canon sensor has been tested in our laboratory. The 10M sensor has similar physical size with the 120M one but with a larger pixel size of 7.2μm. According to Canon, the characteristics of the 120M camera are very similar to the 10M pixel camera.

The 10M pixel camera is equipped with Canon software for the readout and configuration. There is no temperature regulation for the camera so the dark measurement might vary with sensor temperature. However, no dark current was detected at the environment temperature between 20~28 degree C with 2s exposures. Given the exposure time allowed for the metrology camera, we do not expect dark current will contribute excess noise to the images. The readout noise is around 4 e- under the highest gain setting under 10 fps readout. The corresponding dynamic range is about 1500. At lowest gain, the dynamics range can be increase to about 3000. The peak quantum efficiency is about 45% at 500nm. The characteristics of the sensor are very close to the scientific CCDs despite of a lower QE. Since the metrology camera receives the signal from the light source through the fiber, lower QE is not a critical parameter for our application. We expect the 120M pixel CMOS sensor should be suitable for the metrology camera.

One of the major concerns for using front illuminated CMOS sensor is the effects of the insensitive area and the micro-lens in each pixel to the centroid determination accuracy. Also, the pixel to pixel response variation is larger in CMOS sensors. To investigate such effect, a back illuminated e2v 44-20 CCD camera and the 10M pixel Canon CMOS camera were used to compare the centroid measurement accuracy with a well sampled spot. A fiber with focal reducing lens was used to generate the light spot for the measurement. Figure 2 shows the setup of measurement with the CCD camera.

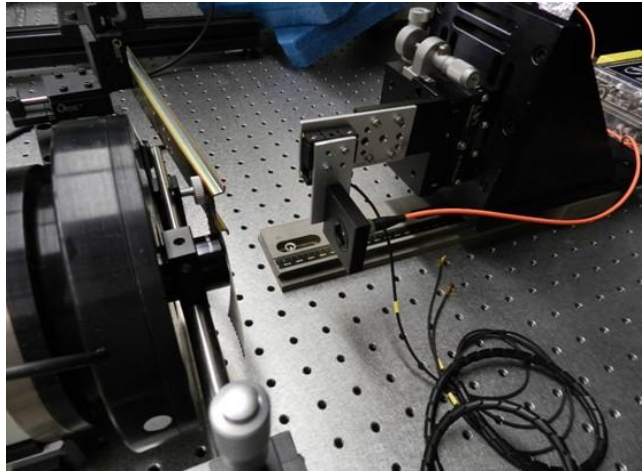

Figure 2. The setup of the centroid test with the CCD camera at the left hand side.

To make reasonable comparison, in both tests, the spot size is adjusted to spread over 5 pixels which is similar to the optical performance of metrology camera. The fiber is driven with a motor drive stage with a repeatability of 2μm. The centroid is calculated by a Gaussian fitting without any image processing. The light source was moved with a step of 2μm for 20 μm. 10 frames were taken at each position to confirm the repeatability of the measurement. For each step, the measurement error of the 10 frames is less than 0.1μm, which shows good repeatability of the setup and the centroid algorithm. Figure 3 shows the average position of the measurement for the CCD camera and the Canon CMOS camera. As we do not know the real position of the light source on the driving motor, the linear fitting of the spots was made assuming the light source moved linearly. The precision of the centroid measurement is defined as the different between the expected position from the linear fitting and the measured position. In both CCD and CMOS cameras, similar results were obtained and the precision of the centroid measurement is within 0.1μm. Similar results were obtained when the image moved more than 20μm. In principle, the measurement accuracy of CMOS sensor should be two times better than the case of CCD. The pixel size is 7.2μm and 15μm for CMOS and CCD sensor respectively. Thus, the performance of CMOS sensor is worse than the performance of CCD. This is probably due to the frontside illuminated structure in CMOS as well as higher pixel to pixel variation in CMOS sensor. However, in both case the results are better than 1/25 of the pixel size (0.6 and 0.288). The performance is good enough for our requirement.

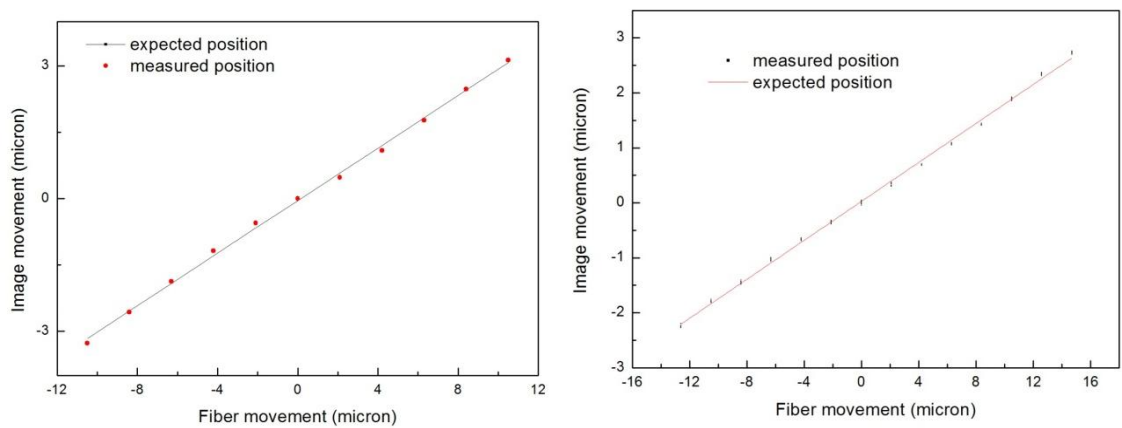

Figure 3. The centroid testing result for CCD camera (left) and Canon 10 M CMOS camera (right).

## 3.2 Camera optics

The optics of metrology camera produces an image of the focal plane with required magnification factor. With the 120M 2.2 μm pixel camera, the magnification factor for the camera optics should be between 0.0176 and 0.0457. To provide better centroid accuracy, the magnification factor should be as large as possible. On the other hand, a large magnification factor will increase the difficulties in the alignment of metrology camera with the focal plane and require a better precision of the optics. A magnification factor around 0.04 might be a sweet point for the metrology camera. Meanwhile, the camera optics should give low field distortion and a uniform point spread function across the field.

The design of optics consists of five spherical lenses with a total length around 645mm as shown in figure 4. The focal length is about 18 m. It is a telecentric design with an angle of only 0.017 degree at the field edge. The aperture of the camera is about 130 mm. Given the possible available aperture size of the optics, the optical performance is diffraction limited across the field. The expected spot size of the fiber tip extends about 10μm or ~5 pixels, even without de-focusing, and satisfies the centroid measurement requirement. The magnification factor is 0.0415 and the distortion at the edge of the field is 0.0005 which is much smaller than the distortion of WFC (0.0097).

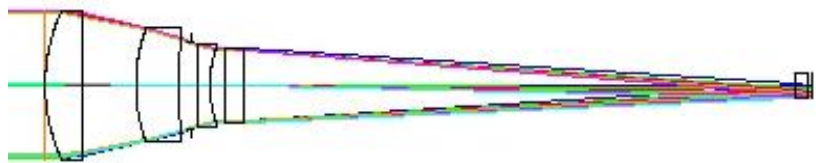

Figure 4. The optical layout of the metrology camera.

As mentioned, the metrology camera will also be used for FMOS. The optical performance of the metrology camera with FMOS WFC is still diffraction limited in the FMOS focal plane range after refocusing of the camera detector. The focus distance change is about 15mm between PFS and FMOS cases.

The lenses are custom components made by a commercial supplier (BASO Precision Optics Ltd.). All lenses are coated with narrow pass band (30nm FWHM) coating to fit the back lit light source and provide the possibility of operating the metrology camera at day time for system calibration.

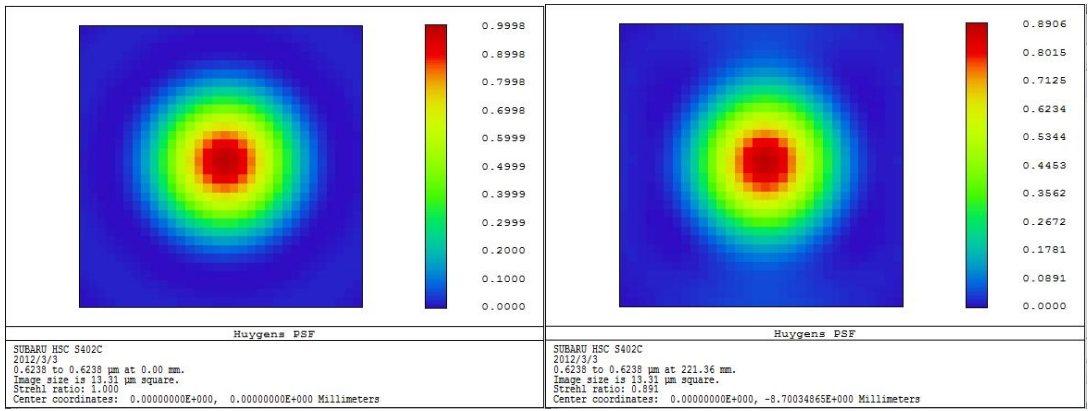

Figure 5. The point spreading function for field center (left) and field edge (right) of the PFS metrology camera.

## 3.3 Camera mechanics

The mechanical system consists of the following elements assuming metrology camera is installed as a Cassegrain instrument:

1) A lens tube that holds the lenses and detector into one piece. Due to the possible thermal expansion, aluminum is not suitable for the base plate and stainless steel is required to prevent large focus adjustment. A focusing stage for the detector will be included for the focus position at different temperatures and also for the different focal positions for PFS and FMOS.

2) A supporting structure with baseplate for the lens tube with initial alignment capability with good precision during the commissioning.

3) A box with mounting structure that could be permanently attached to the telescope structure. This provides the required interface to fit our system to Cassgrain bonette of Subaru telescope so that the system can be positioned with a repeatable accuracy.

The video mode of CMOS camera showing a live image of the back-lit fiducial fibers will be used for the alignment of the camera to the positioners. It is envisioned that once this is done in initial commissioning the camera system can be remounted accurately and will not need to be re-aligned.

### 3.4 Control and operation

The metrology camera calculates the location of the scientific fibers and sends back the position for each scientific fiber for the COBRAs to iterate the fiber positions. The metrology camera system controls the back illuminator so the fibers will be illuminated when the COBRAs reach the next position.

In the beginning of each positioning, the COBRAs will move to the home position to generate the reference points across the field. The fiducial fibers will be also lit up for the reference image and also imaged at the same time for accurate mapping of science fiber locations. The science fibers are then moved to a new position based on the model of the positions for the chosen targets in a new object field. In each sequence, the error from the required positions is calculated and the fibers move to the new positions to reduce the error. This process is then repeated a number of times to converge the fiber on the required positions.

The CMOS sensor is linked to the control computer with CoaXpress interface which allows 3 Gbps data communication and more than 50 meter long cables. The frame rate with 120M pixel camera is 1.5 fps. The raw image data will be stored in the memory of the control computer to allow fast access and image process for the centroid calculation.

The main task of the metrology camera software is to identify the science and fiducial fibers and measure centroids of fiber spots. These centroids along with fiber identifications are then sent to the COBRA control system. The centroiding includes adjusting for pixel sensitivity, bad pixels, and the pixel DC offset, which are pre-established. The camera software receives system commands and information and sends measured centroids to the COBRA control software. The software also controls the back illumination sources so the science fibers are lit on at the right time for the metrology camera. The software also provides diagnostic and test capabilities of the CMOS detectors.

The centroid algorithm is very important to provide solid centroid calculation in a short time for all fibers (< 2s). PFS fitting usually generates better centroid accuracy. However, the fitting might not converge under certain conditions and takes more CPU time. On the other hand, centroid estimation by center of mass is usually very robust and fast but the error might be larger. We are now comparing the two different methods with both simulations and with the dome seeing test images. We plan to implement both algorithms during the testing of metrology camera to examine the robustness and accuracy performance. This will help to decide the final algorithm to be used in the metrology camera system.

## 4. SUMMARY

The concept design of PFS metrology camera is presented. The concept of a Cassgrain metrology camera is proved to be feasible to provide the required accuracy. The optical design has been done and detailed performance and tolerance analysis is being investigated. We expect to get the optics of PFS metrology camera in Oct 2012 following by the CMOS camera delivery. The integration and testing of PFS metrology camera will start by the end of 2012 in Taiwan for basic characteristics. It will be sent to JPL in late 2013 for the testing of COBRA module and the used to final integration of the whole COBRA operation.

## ACKNOWLEDGEMENT

We gratefully acknowledge support from the Funding Program for World-Leading Innovative R&D on Science and Technology(FIRST) "Subaru Measurements of Images and Redshifts (SuMIRe)", CSTP, Japan for PFS project. The work in ASIAA, Taiwan is supported by the Academia Sinica of Taiwan.